\documentclass[twocolumn,fleqn,usenatbib]{aastex631}
\usepackage[T1]{fontenc}
\usepackage{physics}
\usepackage{tabularx}
\usepackage{graphicx}	% Including figure files
\usepackage{amsmath}	% Advanced maths commands
\usepackage{amssymb}	% Extra maths symbols
\usepackage{newtxtext,newtxmath}

\begin{document}

\title{Inferring kilonova ejecta photospheric properties from early blackbody spectra}

\author[0009-0003-0141-6171]{Gilad Sadeh}
\affiliation{Dept. of Particle Phys. \& Astrophys., Weizmann Institute of Science, Rehovot 76100, Israel}

\begin{abstract}
We present simple analytic corrections to the standard blackbody fitting used for early kilonova emission. We consider a spherical, relativistically expanding shell that radiates thermally at a single temperature in its own rest frame. Due to relativistic effects, including Doppler boosting, time delay, and temperature evolution- the observed temperature is smeared across different polar angles by approximately $\sim10\%$. While the observed spectrum remains roughly consistent with a single-temperature blackbody, neglecting relativistic effects leads to significant systematic inaccuracies: the inferred photospheric velocity and temperature are overestimated by up to $\sim50\%$ for mildly relativistic velocities. By applying our analytic corrections, these deviations are reduced to within $10\%$, even in cases where the photosphere is receding and cooling is considered. Applying our corrections to observed kilonovae (AT2017gfo and the thermal component of GRB211211A) reveals that standard blackbody fitting overestimated the inferred velocities and temperatures by $10\%-40\%$, such deviations can alter the inferred formation of heavy elements.

\end{abstract}


%%%%%%%%%%%%%%%%%%%%%%%%%%%%%%%%%%%%%%%%%%%%%%%%%%

%%%%%%%%%%%%%%%%% BODY OF PAPER %%%%%%%%%%%%%%%%%%

\section{Introduction}
The detection of electromagnetic counterparts to gravitational-wave (GW) events has opened a new window into the physics of neutron star mergers. The landmark observation of GW170817 and its associated kilonova, AT2017gfo, provided crucial insights into neutron star coalescence, the origin of heavy elements via rapid neutron capture (\textit{r}-process), and the interaction between relativistic outflows and their environment \citep[e.g.,][]{abbott_gw170817_2017, kasen_origin_2017, pian_spectroscopic_2017, tanvir_emergence_2017}. Kilonovae are powered by the radioactive decay of freshly synthesized heavy elements in the merger ejecta, producing a characteristic thermal emission that evolves from blue to red as the ejecta expands and cools \citep{li_transient_1998}.  
The early-phase kilonova emission carries valuable information about the fastest ejecta components, yet its interpretation remains challenging. The properties of this rapidly expanding material, including its mass, velocity, composition, and heating rate, are key to understanding mass ejection mechanisms and the synthesis of heavy elements during and shortly after the merger \citep[e.g.,][]{metzger_kilonovae_2019, kasliwal_spitzer_2022}. However, theoretical models for early kilonova light curves remain uncertain due to complexities in radiative transfer, ionization balance, opacities, and additional energy sources. These models rely on numerical simulations that incorporate numerous highly uncertain parameters \citep[e.g., composition, radioactive heating, and line structure of high-Z elements,][]{kasen_origin_2017}, further complicating predictions.

The kilonova AT2017gfo, associated with GW170817, remains the most well-characterized event to date. Early optical and ultraviolet observations revealed rapidly evolving blue emission, with initial temperatures exceeding $10000$K and photospheric surface velocities reaching up to $0.3c$ \citep{chornock_electromagnetic_2017,nicholl_electromagnetic_2017,pian_spectroscopic_2017,smartt_kilonova_2017,shappee_early_2017, nicholl_electromagnetic_2017,villar_combined_2017, metzger_welcome_2017, waxman_constraints_2018}. Over the course of approximately a week, the emission gradually transitioned to redder colors. Spectroscopic studies have reported the identification of \textit{r}-process elements such as strontium, reinforcing the role of neutron star mergers as a significant site of heavy element production \citep{watson_identification_2019, sneppen_discovery_2023}. Additionally, an analysis of the observed X-shooter spectra suggests that the ejecta exhibited a highly spherical geometry at early times \citep{sneppen_spherical_2023}.  
Early observations of AT2017gfo were reasonably well described by a single-temperature blackbody spectrum \citep{drout_light_2017, mccully_rapid_2017, pian_spectroscopic_2017, shappee_early_2017, waxman_constraints_2018}. However, \citet{sneppen_blackbody_2023} pointed out that the mildly relativistic velocities of the ejecta are expected to distort the observed spectrum, even if the emission originates from a single-temperature thermal emission. These distortions arise from two key relativistic effects: (i) variations in Doppler boosting across different polar angles (or angular sections), depending on the velocity projection along the observer's line of sight, and (ii) the continuous cooling of the ejecta ($T \propto t^{-0.5}$ for AT2017gfo), combined with significant time-delay effects. The latter implies that angular sections observed at later times (closer to the front) appear cooler than more distant angular sections, which correspond to earlier phases of the evolution.
Beyond GW170817, another kilonova candidate has been proposed in association with GRB 211211A \citep{troja_nearby_2022,rastinejad_kilonova_2022}.  
Future kilonova observations within the first 12 hours post-merger are expected to capture bright blue emission with temperatures exceeding $10,000$K while the photosphere is still within the fast ejecta tail, reaching velocities of $0.6$-$0.7c$ \citep{radice_binary_2018, nedora_dynamical_2021, nedora_numerical_2021, fujibayashi_comprehensive_2023, hajela_evidence_2022, radice_new_2022, rosswog_heavy_2024}. This fast ejecta component is typically studied in the context of its non-thermal emission, expected to emerge on timescales of years \citep{nakar_detectable_2011, kathirgamaraju_observable_2019, nedora_modelling_2023, sadeh_non-thermal_2023, sadeh_non-thermal_2024, sadeh_late-time_2024}.  
However, detecting such early-phase emission poses significant observational challenges, requiring prompt follow-up with wide-field, high-cadence surveys. ULTRASAT, with its high sensitivity in the ultraviolet (UV) and wide-field coverage, is uniquely suited to overcome these limitations \citep{sagiv_science_2014, shvartzvald_ultrasat_2024}.

In this work, we build upon these insights to develop a framework for inferring photospheric properties from the early blackbody spectrum emitted by a relativistic spherical shell. Our goal is to improve our understanding of neutron star merger dynamics and the associated nucleosynthesis. In this analysis, we neglect frequency-dependent effects and assume that all photons originate from the same local temperature and radius. This approximation does not capture the full complexity of radiative transfer in kilonovae, which may involve multi-temperature emitting regions, wavelength-dependent opacities, and line blanketing. While we adopt a single-temperature blackbody model, since the early spectra measured are consistent with a simple blackbody, we acknowledge that such spectra may arise as an emergent coincidence of these effects rather than reflecting a truly uniform temperature structure.  
We examine how the temperature and velocity shape the observed thermal emission during the first hours to days post-merger, and show that the intrinsic photospheric temperature can be reliably estimated. This provides a critical diagnostic for the ejecta’s physical conditions, the heating processes at play, and the validity of local thermodynamic equilibrium. Our findings offer a direct pathway to constraining the merger's early emission properties and improving models of kilonova light curves.
The structure of this paper is as follows: In \S~\ref{sec:calculation}, we present our theoretical framework and detailed calculation of the observed spectrum across a range of velocities and temperatures. \S~\ref{sec:analytic} introduces a simplified analytic approach for inferring the photospheric velocity and temperature from observed spectra. In \S~\ref{sec:existing}, we apply this method to refine constraints on the ejecta photospheric velocity and temperature from previous kilonovae observations. Finally, in \S~\ref{sec:conclusions}, we summarize our findings and discuss their implications.

\section{Blackbody emission of a relativistic spherical shell}
\label{sec:calculation}
\citet{sneppen_blackbody_2023} investigated the observed spectrum of an expanding relativistic spherical shell emitting thermal radiation, specifically analyzing deviations from a pure blackbody spectrum. For mildly relativistic velocities, these deviations are on the order of a few percent, making them difficult to distinguish from other effects, such as wavelength-dependent photospheric variations or significant absorption features, especially when the spectrum is inferred from photometry.  
In this work, we focus on inferring velocity and temperature from the observed blackbody spectrum, rather than characterizing deviations from it.

\subsection{Semi-analytic calculation}
We will follow \citet{sneppen_blackbody_2023} notation with some additions, see Fig. \ref{fig:illustration} for a schematic illustration: 
\begin{itemize}
    \item The photosphere surface is defined with its radial, $R_\text{ph}$, azimuthal, $\phi$, and polar, $\theta$, coordinates. We also use $\mu\equiv\cos(\theta)$.
    \item $\beta_\text{ph}(t)=\beta_0\left(\frac{t}{t_0}\right)^{-k},\gamma(t)$ are the ejecta velocity (in units of $c$) and Lorentz factor at the photosphere, also denoted as $\beta_\text{ph},\gamma_\text{ph}$. $t$ is the time in the lab frame, $t_0$ and $\beta_0$ are arbitrary initial time and velocity.
    \item $T_\text{ph}=T_0\left(\frac{t}{t_0}\right)^{-\alpha}$ it the ejecta rest frame temperature at the photosphere while $T_\text{obs}=\delta_D T_\text{ph}$ is the Doppler shifted temperature (The Doppler correction to a Planck spectrum is obtained by simply shifting the temperature with the Doppler factor) at the observer frame, where $\delta_D\equiv \frac{1}{\gamma_\text{ph}(1-\beta_\text{ph}\mu)}$. $T_0$ is an arbitrary initial temperature. 
    \item The observed time $t_\text{obs}$ of a photon that was emitted from $R_\text{ph}=\int_0^t \beta_\text{ph} cdt'$, and from a polar angle $\theta$, is $t_\text{obs}=t-\frac{R_\text{ph}\mu}{c}$.
\end{itemize}
The emitted luminosity per unit frequency is given by
\begin{equation}
L_\nu=4\pi D_L^2\int_{\Omega_\text{obs}}I_\nu\cos{\theta_\text{obs}}d\Omega_\text{obs},
\end{equation}
where $I_\nu$ is the specific intensity, $\Omega_\text{obs}$ is the observer solid angle, $\theta_\text{obs}$ is observer polar angle, and $D_L$ is the luminosity distance to the emitting object.
\citet{sneppen_blackbody_2023} showed this can be written as 
\begin{equation}
\label{eq:integral}
    L_\nu=8\pi^2\int B_\nu(T_\text{obs}(t(t_\text{obs},\mu),\mu))\left(R_\text{ph}(t(t_\text{obs},\mu))\right)^2\mu d\mu,
\end{equation}
considering cylindrical symmetry, where $B_\nu$ is simply the Planck function. This integral can be calculated numerically for various values of $\beta_0,k,T_0,\alpha$. In all of the calculations, we use $t_0\ll6$ hours. 

An important clarification should be made regarding the integral limits:  
(i) The lower limit corresponds to emission along the observation axis, given by $\mu = 1$ (i.e., $\theta = 0$).  
(ii) The upper limit is determined by the largest polar angle, $\theta$, from which radiation can escape without being absorbed by the optically thick material at $R < R_\text{ph}$. This upper bound is obtained by calculating the maximum perpendicular distance from the line of sight, given by $R_\text{ph}(t(t_\text{obs},\mu))\sin\theta$.  
Numerically, this can be computed by
\begin{equation}
\partial_\mu\left( R_\text{ph}(t(t_\text{obs},\mu))\sqrt{1-\mu^2}\right)=0.
\end{equation}
In the case of constant velocity, $R_\text{ph}(t(t_\text{obs},\mu))=\frac{\beta_\text{ph} c t_\text{obs}}{1-\beta_\text{ph}\mu}$, yielding the following
\begin{equation}
    \partial_\mu\left(\frac{\sqrt{1-\mu^2}}{1-\beta_\text{ph}\mu}\right)=0,
\end{equation}
which provides $\mu=\beta_\text{ph}$. This was shown in a slightly different approach in \citet{sadun_relativistic_1991}.

\subsection{Angular distribution}
\begin{figure}
    \centering
\includegraphics[width=8cm]{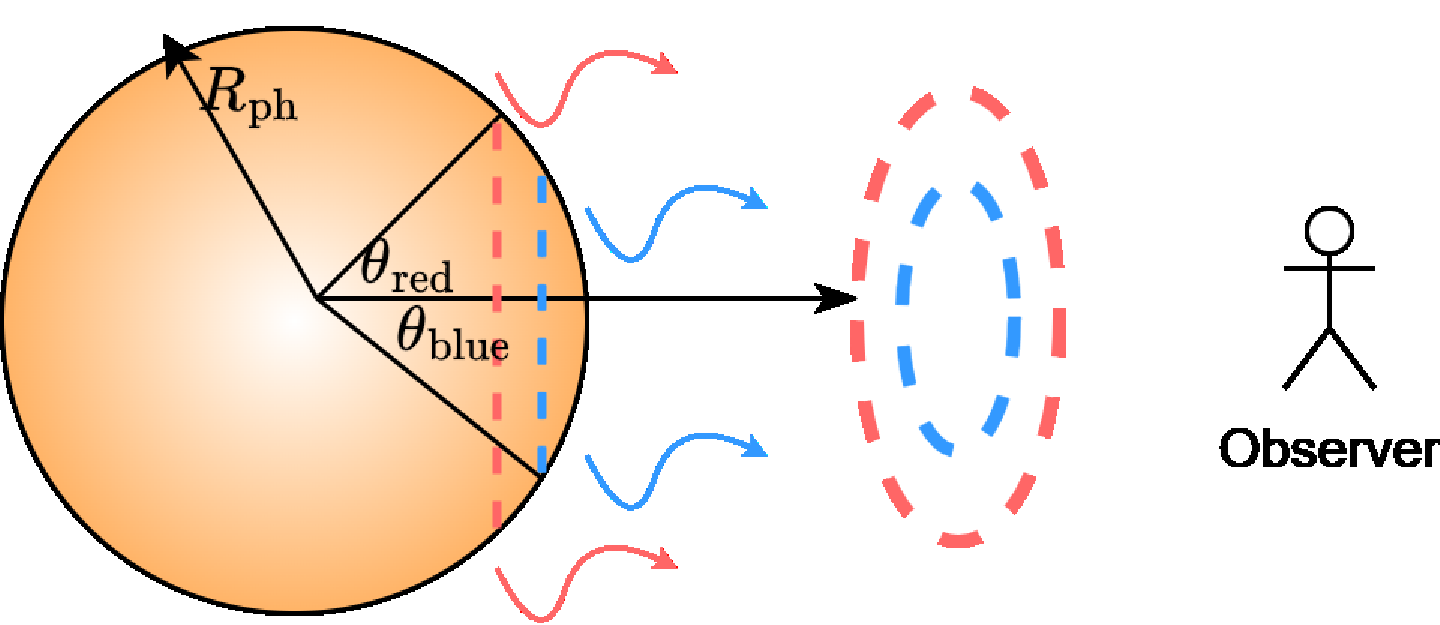}
    \caption{Schematic illustration of the different contributions from different angular sections ("rings") to the observed spectrum. For the same observation time, $t_\text{obs}$, each angular section emits radiation in different lab time, $t$, contributing angular section is boosted with different $\delta_D$, different ring radius, and, in case of cooling, different emitting temperature.}
     \label{fig:illustration}
\end{figure}

The observed spectrum is the sum of contributions from different angular sections, each corresponding to a distinct polar angle (see Fig. \ref{fig:illustration} for a schematic illustration). For a fixed observation time, $t_\text{obs}$, radiation from each angular section is emitted at a different lab-frame time, $t$. Each contributing angular section experiences a different Doppler boost factor, $\delta_D$, a different projected ring radius, and, in the presence of cooling, a different emitting temperature. 

\begin{figure*}
    \centering
    \gridline{
        \fig{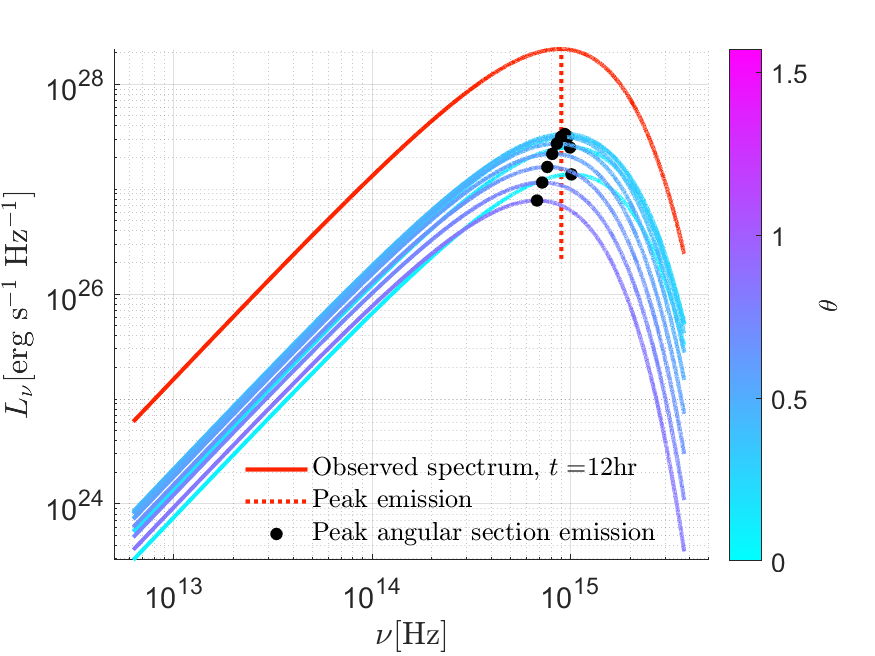}{0.45\textwidth}{(a) }
        \fig{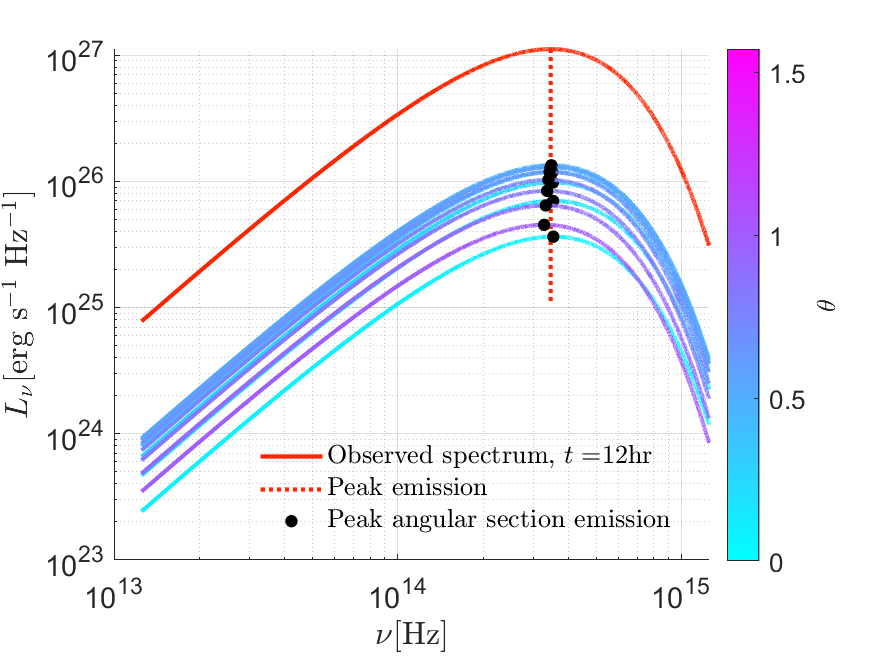}{0.45\textwidth}{(b) }
        }
    \caption{In red: The observed spectrum (with its peak emission) from a mildly relativistic spherical shell emitting thermally with the following parameters $\{\beta_0=0.5,T_0=10^4\text{K}\}$(left panel, constant velocity and temperature)/$\{k=0.3,\alpha=0.5\}$(right panel).
    In light blue to magenta: the decomposition of the spectrum to the contribution from different angular sections. In black: the peak emission from each angular section.}
     \label{fig:angular}
\end{figure*}

In Fig. \ref{fig:angular}, we decompose the observed spectrum into its angular components for both a constant velocity and temperature scenario ($k=0, \alpha=0$) and a receding photosphere with cooling ($k>0, \alpha>0$). In both cases, we find that the observed temperature closely matches the temperature of the angular component with the largest peak emission (i.e., the maximum value of $L_\nu$), while contributions from nearby angular sections introduce minor smearing in opposite directions.  
The temperature spread among the dominant angular components is relatively narrow, approximately $\sim10\%$, and this result remains consistent across various parameter sets, indicating significant limb darkening can be neglected. Since emission from a specific angle originates from a single moment in time, each angle is associated with a well-defined velocity.  
Although \citet{sneppen_blackbody_2023} demonstrated that temperature variations across different polar angles can reach up to $40\%$, in practice, the combined effects of geometric ring area and absorption at large angles reduce this variation to about $10\%$.
An important point to emphasize is that across a wide range of parameters $\{k, \alpha\}$, the observed spectrum from a mildly relativistic spherical shell remains consistent with a simple blackbody spectrum, even in the absence of cooling (see \S  \ref{sec:analytic}).
 
\section{Inference of the photosphere parameters}
\label{sec:analytic}
Regardless of whether cooling or a receding photosphere is considered, the temperature spread among the contributing angular sections remains approximately $10\%$. Consequently, at a given observation time, the observed spectrum from an expanding, relativistic, spherical shell can be well approximated by emission from a single angular section with a single velocity and temperature- specifically, the section with the largest peak emission. This insight allows us to derive an analytic prescription for determining the velocity and temperature of the emitting material.

\subsection{Velocity}
\begin{figure}
    \centering
\includegraphics[width=\columnwidth]{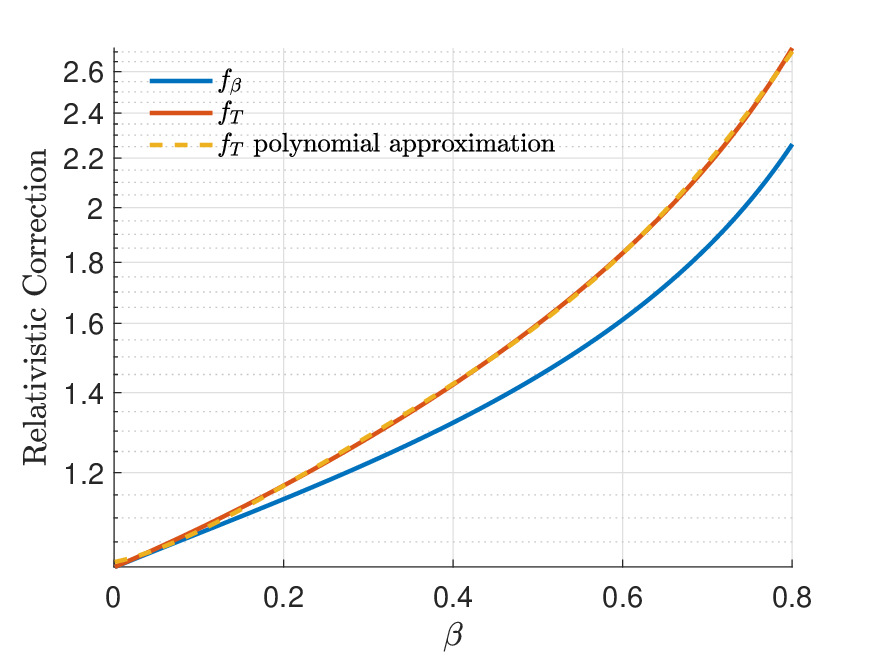}
    \caption{The relativistic corrections to the velocity, $f_\beta$, and temperature, $f_T$, obtained from simple blackbody spectrum fitting. Typically, for mildly relativistic velocities, there is a $\sim50\%$ correction to the velocity and temperature, compared to the non-relativistic treatment.}
     \label{fig:relativistic_corrections}
\end{figure}

For a constant velocity, Eq. (\ref{eq:integral}) simplifies to \citep{sneppen_blackbody_2023}:  
\begin{equation}
\label{eq:const_b}
    L_\nu=8\pi^2(\beta_\text{ph} c t_\text{obs})^2\int_{\beta_\text{ph}}^1  B_\nu(T_\text{obs}(t(t_\text{obs},\mu),\mu))\left(\frac{1}{1-\beta_\text{ph}\mu}\right)^2\mu d\mu.
\end{equation}
\citet{sneppen_blackbody_2023} found that for mildly relativistic velocities \textbf{($\beta\lesssim0.5$)}, deviations from a simple blackbody fit, in this case, are only a few percent. As a result, Eq. (\ref{eq:const_b}) can be approximated as:
\begin{equation}
\begin{aligned}
\label{eq:bb}
    L_\nu&\approx 8\pi^2(\beta_\text{ph} c t_\text{obs})^2 B_\nu(T_\text{fit})\int_{\beta_\text{ph}}^1  \left(\frac{1}{1-\beta_\text{ph}\mu}\right)^2\mu d\mu,\\
    &\equiv4\pi^2(\beta_\text{fit} c t_\text{obs})^2B_\nu(T_\text{fit}),
\end{aligned}
\end{equation}
where $T_\text{fit}$ is a fitted temperature, $\beta_\text{fit}\equiv f_\beta(\beta_\text{ph})\times\beta_\text{ph}$, and $f_\beta(\beta)$ is defined as:
\begin{equation}
    f_\beta(\beta)\equiv \sqrt{\frac{2}{\beta^2}\left(\ln(1-\beta)-\ln(1-\beta^2)+\frac{\beta}{(1-\beta^2)}\right)}.
\end{equation}
In the limit $\beta\rightarrow0$, $f_\beta(\beta)\rightarrow1$, and Eq. (\ref{eq:bb}) reduces to the standard non-relativistic blackbody emission from a sphere. For mildly relativistic velocities, however, there is typically a $\sim50\%$ correction to the velocity compared to the non-relativistic treatment (see Fig. \ref{fig:relativistic_corrections}).

\begin{figure*}
    \centering
    \gridline{
        \fig{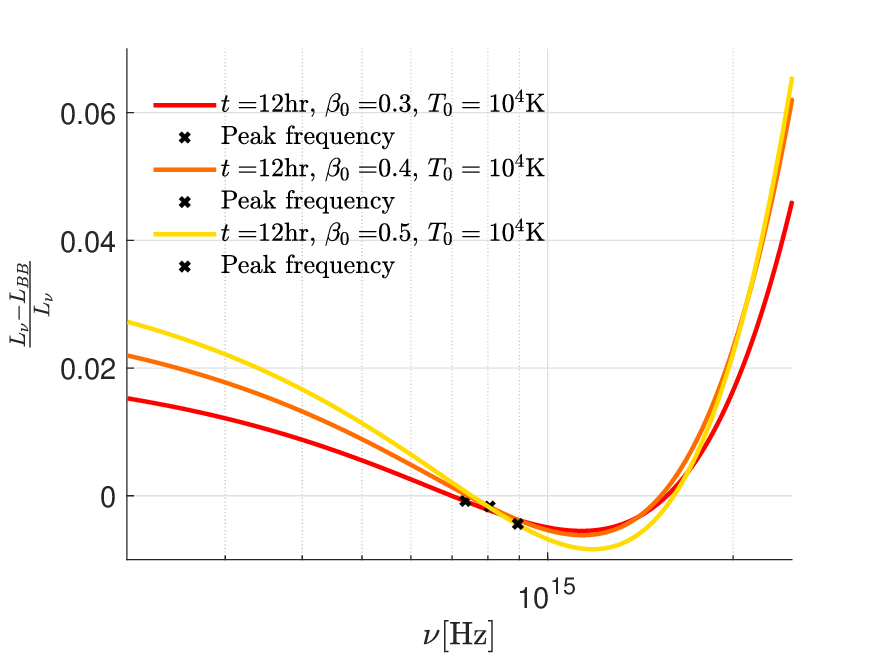}{0.45\textwidth}{(a) }
        \fig{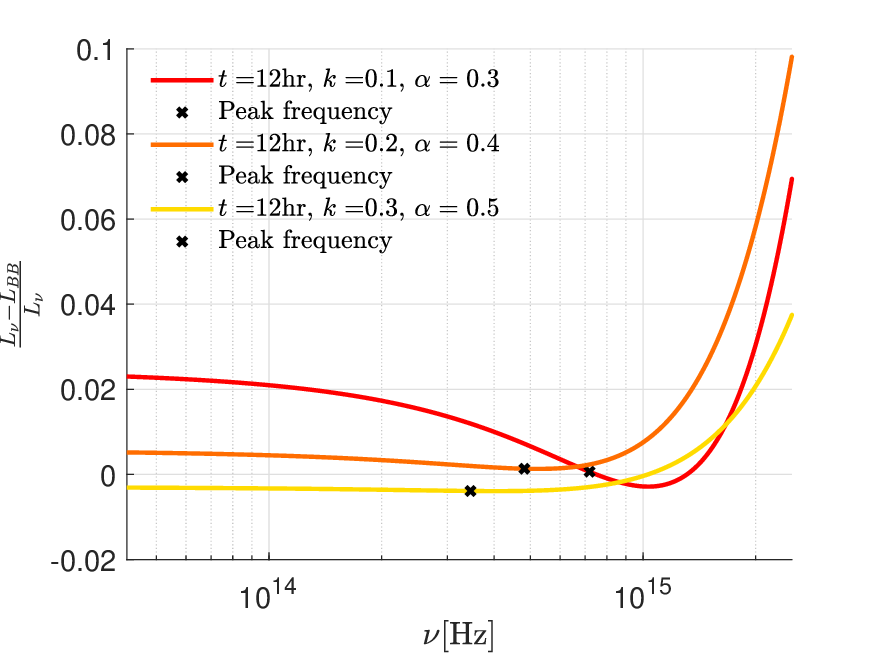}{0.45\textwidth}{(b) }
        }
    \caption{The normalized difference between the full calculation of Eq. (\ref{eq:integral}), $L_\nu$, and the best-fit blackbody spectrum using Eq. (\ref{eq:bb}), $L_\text{BB}$, for various parameter sets $\{\beta_0, k, T_0, \alpha\}$ (left panel- constant velocity and temperature, right panel- decelerating velocities and temperatures). The fitting procedure adjusts both the temperature and normalization. The observed spectrum is well described by a single blackbody for both decelerating and constant velocity and temperature scenarios (less than $10\%$ deviations for all fits and considered frequencies). The inferred velocity, $\beta_\text{ph}$, closely matches the actual velocity from the calculations (Table \ref{tab:cases1}).}
     \label{fig:fitting}
\end{figure*}

In Fig. \ref{fig:fitting}, we fit Eq. (\ref{eq:bb}), adjusting both temperature and normalization to the full calculation from Eq. (\ref{eq:integral}) across various parameter sets, $\{\beta_0, k, T_0, \alpha\}$. We find that the observed spectrum is well described by a single blackbody, regardless of whether the velocity and temperature are constant or decelerating, as long as the velocity remains below $\beta < 0.8$. Furthermore, the inferred velocity, $\beta_\text{ph}$, closely matches the actual velocity used in the calculations (see Tables \ref{tab:cases1},\ref{tab:cases2}).

\subsection{Temperature}
In the previous subsection, we demonstrated that within the expected parameter space, a simple single-velocity fit to the observed blackbody spectrum provides estimates for $T_\text{fit}$ and $\beta_\text{ph}$. However, since the Doppler boost, $\delta_D \equiv \frac{1}{\gamma_\text{ph}(1-\beta_\text{ph}\mu)}$, is angle-dependent, we must select a specific angle to relate $T_\text{fit}$ to the actual rest-frame temperature of the emitting plasma, $T_\text{ph}$. 
In Fig. \ref{fig:angular}, we show that the angular section with the largest peak luminosity has the same temperature as the entire spectrum. The peak luminosity is obtained at a frequency of $h\nu_\text{peak}=2.82k_BT$ \citep{rybicki_radiative_1979}, and is proportional to:
\begin{equation}
    B_{\nu=\nu_\text{peak}}\propto \frac{\nu_\text{peak}^3}{e^\frac{h\nu_\text{peak}}{k_BT}-1}=\frac{\nu_\text{peak}^3}{e^{2.82}-1}.
\end{equation}
The spectral contribution from different angular sections at the peak is given by:
\begin{equation}
\begin{aligned}
    \frac{dL_{\nu=\nu_\text{peak}}}{d\theta}&=8\pi^2(\beta_\text{ph} c t_\text{obs})^2 B_{\nu=\nu_\text{peak}}(T_\text{fit})  \left(\frac{1}{1-\beta_\text{ph}\mu}\right)^2\mu\sqrt{1-\mu^2},\\
    &\propto\left(\frac{1}{1-\beta_\text{ph}\mu}\right)^3  \left(\frac{1}{1-\beta_\text{ph}\mu}\right)^2\mu\sqrt{1-\mu^2}.
    \end{aligned}
\end{equation}
where the $\sqrt{1-\mu^2}$ factor arises from the Jacobian.  
As shown in Fig. \ref{fig:angular}, the best-fit temperature corresponds to the angular section that maximizes $\frac{dL_{\nu=\nu_\text{peak}}}{d\theta}$:
\begin{equation}
\frac{-3\beta_\text{ph}\mu_\text{max}^3-2\mu_\text{max}^2+4\beta_\text{ph}\mu_\text{max}+1}{\left(-\beta_\text{ph}\mu_\text{max}+1\right)^6\sqrt{-\mu_\text{max}^2+1}}=0.
\end{equation}
This equation can be solved numerically to obtain $\mu_\text{max}(\beta)$. The relation between the fitted temperature and the actual photospheric temperature in the ejecta rest frame is then:
\begin{equation}
    T_\text{fit}=\frac{T_\text{ph}\sqrt{1-\beta_\text{ph}^2}}{(1-\beta_\text{ph}\mu_\text{max})} \equiv f_T(\beta_\text{ph})\times T_\text{ph},
\end{equation}
where $0.75\lesssim\mu_\text{max}\lesssim0.98$ for $0.1<\beta_\text{ph}<0.8$.
For $\beta \rightarrow 0$, $f_T(\beta) \rightarrow 1$, ensuring that the fitted temperature converges to the actual temperature of the emitting matter at the photosphere in the non-relativistic limit. However, for mildly relativistic velocities, a $\sim50\%$ correction to the temperature is required compared to the non-relativistic case (see Fig. \ref{fig:relativistic_corrections}).  
To approximate $f_T(\beta)$ within $1\%$ accuracy for $\beta < 0.8$, we use a fourth-degree polynomial:
\begin{equation}
    f_T(\beta)\approx 1.01+0.35\beta+3.32\beta^2-6.64\beta^3+6.56\beta^4.
\end{equation}
In conclusion, to properly fit and infer the velocity and temperature of a mildly relativistic, expanding spherical shell, the observed spectrum should be fitted using the following functional form:
\begin{equation}
\label{eq:final}
    L_\nu\approx 4\pi^2(f_\beta(\beta_\text{ph})\times\beta_\text{ph} c t_\text{obs})^2B_\nu(f_T(\beta_\text{ph})\times T_\text{ph}).
\end{equation}
\subsection{Parameters inference accuracy}
We verify the accuracy of our simple analytic corrections both for constant velocities and temperatures ($k,\alpha=0$) and for varying ones ($k,\alpha>0$) representative of values expected in kilonovae, constrained by the early temperature and velocity inferred from AT2017gfo \citep[the inferred velocity and temperature at $\sim1$day are $\sim0.25$c and $\sim6500$K, accordingly, along with $k=0.4$ and $\alpha=0.55$,][]{waxman_constraints_2018}. The typical maximal velocity of the ejected matter in binary neutron star mergers simulation is $\beta\approx0.8$ \citep{radice_binary_2018,nedora_dynamical_2021,nedora_numerical_2021,fujibayashi_comprehensive_2023,hajela_evidence_2022,radice_new_2022,rosswog_heavy_2024}. Additionally, with ULTRASAT \citep{sagiv_science_2014,shvartzvald_ultrasat_2024}, we expect to obtain kilonova UV signal on a time scale of $\sim$hours. 
Thus, for varying velocity and temperature ($k>0,\alpha>0$), we obtain few discrete values of $\beta^i_\text{ph}$ and $T^i_\text{ph}$ in few observed times, $t^i_\text{obs}\geq6$hours. Then, by considering a vector of lab times $t^i=\frac{t^i_\text{obs}}{1-\beta^i_\text{ph}\mu^i_\text{max}}$ we fit the following functional form
\begin{equation}
\beta^i_\text{ph}=\beta_0\left(\frac{t^i}{t_0}\right)^{-k}, \quad T^i_\text{ph}=T_0\left(\frac{t^i}{t_0}\right)^{-\alpha},
\end{equation}
to infer $k$ and $\alpha$. We find that the inferred parameters are accurate to a level of $\sim10\%$, see Tables \ref{tab:cases1} and \ref{tab:cases2}. 
\begin{table}
\begin{tabular}{|c|c|}
\hline
Cases $\{\beta_0,T_0\}$ & Inferred values \\
\hline
$\{0.3,10^4\text{K}\}$ & $\{0.30,9.8\times10^3\text{K}\}$ \\
\hline
$\{0.4,10^4\text{K}\}$ & $\{0.39,9.7\times10^3\text{K}\}$ \\
\hline
$\{0.5,10^4\text{K}\}$ & $\{0.49,9.6\times10^3\text{K}\}$ \\
\hline
$\{0.6,10^4\text{K}\}$ & $\{0.58,9.6\times10^3\text{K}\}$ \\
\hline
$\{0.7,10^4\text{K}\}$ & $\{0.69,9.5\times10^3\text{K}\}$ \\
\hline
\end{tabular}
\caption{Inferred values, constant velocity and temperature.}
\label{tab:cases1}
\end{table}
\begin{table}
\begin{tabular}{|c|c|}
\hline
Cases $\{k,\alpha\}$ & Inferred values \\
\hline
$\{0.1,0.3\}$ & $\{0.10,0.28\}$ \\
\hline
$\{0.2,0.35\}$ & $\{0.20,0.31\}$ \\
\hline
$\{0.25,0.4\}$ & $\{0.24,0.36\}$\\
\hline
$\{0.3,0.45\}$ & $\{0.29,0.4\}$\\
\hline
$\{0.35,0.5\}$ & $\{0.34,0.45\}$\\
\hline
$\{0.4,0.55\}$ & $\{0.37,0.5\}$\\
\hline
\end{tabular}
\caption{Inferred values, varying velocity and temperature.}
\label{tab:cases2}
\end{table}
The temperature inferred from fitting the spectrum is also sensitive to the assumed reddening correction. Typical uncertainties in $E(B-V)$ of $\pm0.05$ can propagate to uncertainties in $T_\text{fit}$ of up to $\sim500-1000$K, depending on the spectral slope and wavelength coverage. While we focus on systematic corrections due to relativistic effects, these reddening-related uncertainties should also be considered when interpreting fitted temperatures. We also note that the inferred (photospheric) velocity depends linearly on the assumed distance. As the distance to both AT2017gfo and GRB 211211A is inferred from the redshift of nearby galaxies using $H_0$, the $\sim10\%$ uncertainty in $H_0$ translates directly into a $\sim10\%$ systematic uncertainty in the inferred velocity. This limitation is inherent even when relativistic corrections are applied.
\section{Observed Kilonovae}
\label{sec:existing}
We fit our analytic formula (Eq. (\ref{eq:final})) to the early observations of AT2017gfo \citep{arcavi_optical_2017,nicholl_electromagnetic_2017,waxman_constraints_2018} and to the thermal emission associated with GRB 211211A \citep{troja_nearby_2022,rastinejad_kilonova_2022}, adopting the luminosity values provided within.  
An important note should be made regarding GRB 211211A: while a thermal component has been identified, it is also consistent with thermal emission from dust heated by UV and soft X-ray radiation. This radiation could have been produced by the interaction of the GRB jet plasma with the circumstellar medium \citep{waxman_strong_2025}.

\textbf{GW170817}: AT2017gfo was found to be consistent with a blackbody spectrum during the first few days of observations \citep[in both photometry and spectroscopy,][]{arcavi_optical_2017,nicholl_electromagnetic_2017,waxman_constraints_2018,sneppen_blackbody_2023}. In Table \ref{tab:170817}, we present the corrections to the inferred velocities and temperatures. These corrections are of the order of $10\%-20\%$ since $\beta < 0.3$.  
\begin{table}
\begin{tabular}{|c|c|c|}
\hline
Observed time & Previous fit ($\beta_\text{fit},T_\text{fit}$)& New fit ($\beta_\text{ph},T_\text{ph}$)\\
\hline
$0.5$ days & $0.29,10.3\times10^{3}\text{K}$ & $0.24,8.4\times10^{3}\text{K}$ \\
\hline
$0.6$ days & $0.23,10.8\times10^{3}\text{K}$ & $0.2,9.2\times10^{3}\text{K}$\\
\hline
$0.8$ days & $0.29,7.1\times10^{3}\text{K}$ & $0.25,5.8\times10^{3}\text{K}$\\
\hline
$1$ days & $0.26,6.4\times10^{3}\text{K}$ & $0.22,5.3\times10^{3}\text{K}$\\
\hline
$1.2$ days & $0.24,5.7\times10^{3}\text{K}$ & $0.21,4.8\times10^{3}\text{K}$\\
\hline
$1.5$ days & $0.23,5\times10^{3}\text{K}$ & $0.2,4.2\times10^{3}\text{K}$\\
\hline
\end{tabular}
\caption{AT2017gfo.}
\label{tab:170817}
\end{table}

\textbf{GRB 211211A}: \citet{troja_nearby_2022,rastinejad_kilonova_2022} fitted a blackbody spectrum to the observed emission at three different time epochs and found mildly relativistic velocities. In Table \ref{tab:211211A}, we present the corrections to the inferred velocities and temperatures. As expected, the early-time correction is on the order of $\sim30\%$. While our relativistic corrections reduce the inferred temperature and velocity by up to $40\%$, the full uncertainty budget also includes additional $\sim10\%$ velocity error due to distance ($H_0$) uncertainty and $\sim1000$K temperature error from $E(B-V)$ uncertainty, particularly at early times when UV flux dominates.

\begin{table}
\begin{tabular}{|c|c|c|}
\hline
Observed time & Previous fit ($\beta_\text{fit},T_\text{fit}$)& New fit ($\beta_\text{ph},T_\text{ph}$)\\
\hline
$5$ hours & $0.52,16\times10^{3}\text{K}$ & $0.39,11.3\times10^{3}\text{K}$ \\
\hline
$10$ hours & $0.74,8\times10^{3}\text{K}$ & $0.51,5\times10^{3}\text{K}$\\
\hline
$1.4$ days & $0.25,4.9\times10^{3}\text{K}$ & $0.22,4.1\times10^{3}\text{K}$\\
\hline
\end{tabular}
\caption{The thermal component associated with GRB211211A.}
\label{tab:211211A}
\end{table}

\section{Conclusions}
\label{sec:conclusions}
In this work, we have developed an analytic framework for inferring the photospheric velocity and temperature of relativistically expanding spherical ejecta from its observed blackbody spectrum. Our approach accounts for relativistic effects such as Doppler boosting and time delays, along with variations in temperature and velocity, which, if not properly considered, can introduce systematic deviations in the inferred parameters.
Despite the relativistic nature of the expanding shell, we have shown that the observed spectrum remains well approximated by a single-temperature blackbody. The effective temperature inferred from standard blackbody fitting closely matches the temperature of the angular section with the highest peak emission, with a typical spread of $\sim 10\%$ in temperature, ensuring that the observed spectrum does not significantly deviate from a pure blackbody. By introducing correction factors, $f_\beta(\beta)$ and $f_T(\beta)$ (Fig. \ref{fig:relativistic_corrections}), we provide a method to systematically adjust the fitted velocity and temperature, significantly improving their accuracy for mildly relativistic ejecta.

Applying our method to observations of AT2017gfo and the thermal component of GRB 211211A, we found that standard blackbody fitting overestimates the velocity and temperature by $10\%-40\%$, depending on the relativistic expansion velocity. A temperature difference of $\sim2000$K at early times can shift the dominant ionization state of lanthanide-rich ejecta, altering both radiative cooling and line formation \citep[e.g.,][]{barnes_effect_2013,domoto_lanthanide_2022,sneppen_rapid_2024}. Similarly, inferred expansion velocities impact kinetic energy estimates by $\propto\beta^2$, and thus even $10\%$ corrections translate to $\sim20\%$ shifts in ejecta energy.

The analytical framework developed in this paper is applicable to any thermally emitting, quasi-spherical ejecta propagating at mildly relativistic velocities. This includes, for instance, Fast Blue Optical Transients \citep[FBOTS, ][]{perley_fast_2019, margutti_embedded_2019} and emission from hot cocoons \citep{gottlieb_cocoon_2018,gottlieb_hours-long_2023}.

Finally, additional observational constraints, such as spectral line identification, could further refine the velocity estimates of the ejecta. Combined with the analysis presented in this paper, this approach provides a useful method for testing the sphericity of the ejecta.

\section*{Acknowledgments}
We thank Liv Shalom for her helpful contribution. 

\section*{Data Availability}
This research did not generate any new observational data. All analytical results are derived from the equations provided in the manuscript, and no proprietary software was used.

%\appendix

\bibliography{references} 

\bibliographystyle{aasjournal}

\end{document}